\documentclass{iopart}

\usepackage{color}
\usepackage{psfrag}
\usepackage{graphicx}
\usepackage{multirow}
\usepackage{amssymb,amsfonts,latexsym}
\newcommand{\un}[1]{\textrm{\,#1}}
\usepackage{url}

\makeatletter
\def\url@leostyle{
  \@ifundefined{selectfont}{\def\UrlFont{\sf}}{\def\UrlFont{\small\ttfamily}}}
\makeatother
\begin{document}


\title[Searching for NR signals of black hole binaries with
phenomenological templates]{Searching for numerically-simulated signals
  of black hole binaries with a phenomenological template family}

\author{Luc\'ia Santamar\'ia$^1$, Badri Krishnan$^1$ and John T Whelan$^{1,2}$} 

\address{$^1$ Max-Planck-Institut f\"{u}r Gravitationsphysik
  (Albert-Einstein-Institut), Am M\"{u}hlenberg 1, D-14476 Potsdam, Germany}
\address{$^2$ Center for Computational Relativity and Gravitation
  and School of Mathematical Sciences, Rochester Institute of Technology,
  85 Lomb Memorial Drive, Rochester, NY 14623, USA}

\ead{\mailto{lucia.santamaria@aei.mpg.de},\mailto{badri.krishnan@aei.mpg.de},\mailto{john.whelan@ligo.org}}

\begin{abstract}
  Recent progress in numerical relativity now allows computation of
  the binary
  black hole merger, whereas post-Newtonian and perturbative
  techniques can be  
  used to model the inspiral and ringdown phases. So far, most
  gravitational-wave searches have made use of various
  post-Newtonian-inspired templates to search for signals arising from
  the coalescence of compact binary objects. Ajith \etal have 
  produced hybrid waveforms for non-spinning binary black-hole systems
  which include the
  three stages of the coalescence 
  process, and constructed from them
  phenomenological templates which capture the features of these 
  waveforms in a parametrized form. As a first
  step towards extending the
  present inspiral searches to higher-mass binary black-hole systems,
  we have 
  used these phenomenological waveforms in a search for
  numerically-simulated signals injected into synthetic LIGO data as
  part of the NINJA project. 
\end{abstract}


\section{Introduction}
\label{sec:intro}

An international network of ground-based gravitational-wave (GW)
detectors (LIGO~\cite{Abbott:2007kv}, Virgo~\cite{Acernese2006},
GEO600~\cite{GEOStatus:2006}) operating roughly in the
$10^1-10^4\un{Hz}$ 
band has recently finished taking data at
or close to design sensitivity, and
the world-wide LIGO Scientific Collaboration (LSC) and Virgo  
Collaboration are currently 
involved in analyzing it. Furthermore, space-based
instruments, such as the Laser Interferometer Space Antenna 
(LISA)~\cite{LISA1,Danzmann:2003tv}, are also 
expected to probe a significantly lower-frequency stretch of the GW
spectrum, namely the $10^{-4}-10^{-1}\un{Hz}$ band.
One of the most promising candidate sources for
detection is a compact binary system, composed of either neutron stars
or black holes.  GW emission from such systems, in
particular from those consisting of binary black holes (BBH), is expected
to be detected by these instruments in the near 
future, giving rise to the new field of gravitational-wave astronomy. 

The detection of gravitational signals arising from the coalescence of a
BBH system, as well as the estimation of its physical parameters,
relies heavily on the ability to accurately
model the waveforms emitted during the inspiral,
merger and ringdown stages of the process. Until recently, the usual
approach to the problem consisted entirely of making use of analytical
approximations to the full theory of general relativity. Thus, a 
post-Newtonian (PN) expansion would be used to solve Einstein
equations in 
the initial inspiral part of the coalescence (see~\cite{lrr-2006-4}
and references therein), while perturbation
theory would be
applied to the final ringdown after the two black holes have
merged together. The intermediate phase corresponding to the strongest
gravitational signal remained unmodelled until the breakthroughs occurred
in 2005~\cite{Pretorius:2005gq,Campanelli:2005dd,Baker05a}, when 
several numerical-relativity (NR) groups finally succeeded 
in using their numerical codes to solve for the merger and ringdown
phases of the BBH coalescence, calculating the gravitational waves
associated to the process. Steady progress quickly followed those
pioneering results, leading to a thriving era in the field.

Among the most suitable data-analysis (DA) methods currently used to
look for signals from compact 
binary coalescences in the output data of GW 
interferometers 
is the matched-filter search technique against a given class of
template waveforms. So far, most current LSC searches for compact
binaries in 
the LIGO/Virgo data have made use of different flavours of analytical 
PN templates to search for gravitational
waves~\cite{Abbott:2007xi,Abbott:2009tt}. Further 
efforts have been
undertaken in the recent past to incorporate the full
inspiral-merger-ringdown (IMR) waveforms
with information from NR results in the
different DA packages into the LSC Algorithm
Library~\cite{LAL}. These include the EOBNR 
waveforms based on the effective-one-body
approach~\cite{Buonanno:2007pf} 
and the phenomenological waveforms resulting from hybrid PN-NR
match~\cite{Ajith:2007qp,Ajith:2007kx,Ajith:2007xh}
among others. 

In order to be able to make confident statements about the performance of
diverse DA pipelines and their ability to find
true gravitational waves, data
analysts have developed several techniques, one of which consists of studying
the pipeline's response to signals introduced in the detector strain as
\textit{software injections}. This procedure helps understand the
background noise
present in the detector data and provides insight for fine-tuning the numerous
parameters and thresholds in the different analyses. 
Since the strongest burst of gravitational radiation
originates during the BBH merger, a good
test to check the efficiency of existing DA techniques is to
inject NR signals that contain the final part of the
coalescence into simulated LIGO/Virgo data and proceed with the
GW searches. 

\begin{figure}
  \centering{
  \includegraphics[width=0.7\textwidth]{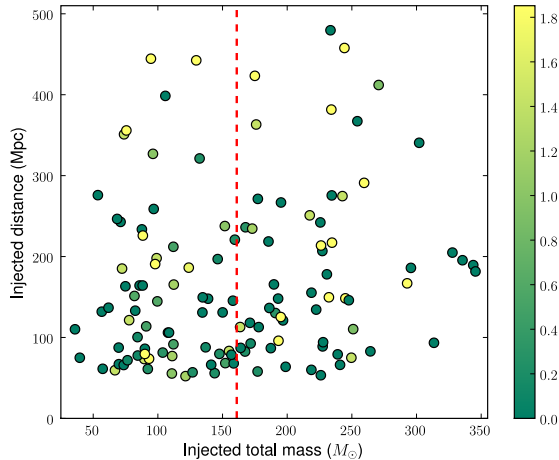}}
  \caption{Total mass and distance of the NINJA injected
    signals, with the colour code encoding the modulus of the
    dimensionless total
    spin $|\vec{S_1}/m_1^2+\vec{S_2}/m_2^2|$ of the black holes. The total
    mass of the injected signals lies within the range 
    $36M_\odot\,\le\,M\,\le460M_\odot$ and they are located
    at a distance between $52$ and $480\un{Mpc}$. The vertical
    line delimits the mass parameter space with $M<160M_\odot$ that
    our phenomenological template bank covers. The full NINJA data
    set spans a duration of a little over 30 hours and contains total
    of 126 signals injected in simulated noise, 67 of which
    overlap with the parameter space of our template bank.}
  \label{fig:ninjaInjPars}
\end{figure}

An open collaboration under the acronym NINJA~\cite{NINJA_web} has been
created, that aims at fostering closer interaction between numerical
relativists and data analysts. Its immediate purpose is to understand
the  
sensitivity of current DA pipelines to binary black hole NR signals
buried in simulated noise. To that effect, a set of waveforms was
injected into LIGO- and Virgo-like noise, within the mass and
distance ranges displayed in Figure~\ref{fig:ninjaInjPars}. 
For the numerical waveforms that were used 
see~\cite{Brugmann:2008zz,Husa:2007hp,
Alcubierre:2000xu,Alcubierre:2002kk,Koppitz:2007ev,Pollney:2007ss,
Imbiriba:2004tp,vanMeter:2006vi,
Zlochower:2005bj,Campanelli:2005dd,
Sperhake:2006cy,
Hinder:2007qu,
Pretorius:2004jg,Pretorius:2005gq,
Scheel:2006gg,
Etienne:2007hr
}, for descriptions of the numerical codes 
see~\cite{Hannam:2007ik,Hannam:2007wf,
Tichy:2008du,
Pollney:2007ss,Rezzolla:2007xa,
Vaishnav:2007nm,Hinder:2007qu,
Buonanno:2006ui,Pretorius:2007jn,
Boyle:2007ft,Scheel:2008rj,
Etienne:2007hr
}.
Several data analysis groups took part in the NINJA project, employing
their data-analysis
codes to search for the NR signals. A comprehensive
compilation  of the goals and results
arising from the NINJA collaboration can be found in a dedicated 
paper~\cite{Aylott08}.

This paper presents the results of a search for BBH signals 
in the NINJA data,
employing a new template bank based on a phenomenological model for 
non-spinning BBH systems by Ajith 
\etal~\cite{Ajith:2007qp,Ajith:2007kx,Ajith:2007xh}. 
These waveforms fully
describe the three stages of the coalescence of two non-spinning black 
holes for 
small mass ratios and, if implemented in current GW searches, should
represent a 
step forward with respect to 
PN templates that do not account for the merger and ringdown 
phases of the process. The remainder of the paper is organized as follows. In 
\Sref{sec:pipeline} we
introduce the data-analysis pipeline developed by the Compact Binary
Coalescence (CBC) group of the LSC to
search for signals from compact binaries in the LIGO/Virgo data. In
\Sref{sec:phenom} we describe the modifications performed in
the pipeline that allow the use of the phenomenological waveforms as a
new search template. In
\Sref{sec:results} we present the results obtained when
applying this new template family to a search in the NINJA
data. Finally in \Sref{sec:conclusion} we review our main
conclusions and propose future improvements and directions for further
research.

\section{The inspiral pipeline}
\label{sec:pipeline}

The LSC inspiral pipeline
infrastructure developed by the CBC group has
been employed to 
analyze the data released for 
the NINJA project. The pipeline, without major conceptual
modifications, 
has been used in LSC searches for compact binaries from the third LIGO
science run onward~\cite{Abbott:2007xi,Abbott:2009tt}. 
The same pipeline has been modified for the analysis of
the NINJA data with the phenomenological template family
described in Section~\ref{sec:phenom}.
Since the signals present in the simulated noise 
are known to be NR simulations of BBH coalescences,
the search method consists of a matched-filter
technique~\cite{Allen:2005fk} using an IMR waveform model
based on the hybrid NR-PN waveforms as described below. The findings 
presented here concentrate solely on the simulated LIGO detectors H1, 
H2 and L1, although the NINJA data was generated for the simulated Virgo
interferometer V1 as well.

The LSC inspiral pipeline performs a series of hierarchical 
operations in order to search for real signals buried in the detector 
noise. The first stage is the construction of a suitable template bank
covering the desired parameter space and for which the overlap between 
any point in the template space
and the nearest template is at least as large as a certain minimal 
match~\cite{Babak:2006,Cokelaer:2007kx}, set as 0.99 in
our search. 
After a filtering process through the desired template bank, the 
triggers that pass a
certain signal-to-noise (SNR) ratio threshold are checked for
coincidence in time
and masses between two or three detectors. For our analyses the
commonly-used threshold value of 5.5 was employed. Whenever triggers 
are found
with comparable coalescence time and parameters (in this case,
component masses), they are stored as
coincident~\cite{Robinson:2008un}. The detection statistic is the combined SNR of the single
detector triggers.

Once the initial matched filter has produced a list of triggers that 
pass the first coincidence step, a second stage follows where data is
again filtered, but only through the templates that previously matched a
trigger. Additionally the $\chi^2$~\cite{Allen:2004gu} and 
$r^2$~\cite{Rodriguez:2007} signal-based vetoes, 
designed to separate
true inspiral signals from fluctuations in non-stationary noise, are applied. 
At this point an 
{\it effective SNR} $\rho_{\mathrm{eff},i}(\rho_i,\chi^2_i)$
is calculated combining the standard SNR with the $\chi^2$
value characterizing the mismatch between the spectral content of
the template and the data. After further coincidence tests,
the surviving
triggers are listed as true gravitational wave
candidates and constitute the output of the search. The significance of
the triggers is based on the combined effective SNR, namely
\begin{equation}
\label{comb_eff_snr} \rho_{\mathrm{eff}} = \sqrt{\sum_{i}^N (\rho_{\mathrm{eff}, i})^2}.
\end{equation}

A direct
comparison between the list of injections performed on the NINJA noise
and the triggers found by the pipeline allows for conclusions about the
sensitivity of the analysis and
the relative performance of the different template banks.

\section{The phenomenological template bank}
\label{sec:phenom}

Most standard searches for gravitational waves from BBHs use the
PN approximation of general relativity to construct banks
of templates that account for the inspiral stage of the coalescence
process, and the final ringdown can also be computed via perturbative
techniques. However, the full calculation of the waveform in the merger stage 
requires numerical methods. These numerical simulations are
in general rather expensive, and it is at present not feasible to model a
coalescing binary over hundreds of orbits with sufficient accuracy.
It is in fact also unnecessary to do so, because PN
theory provides a valid description of the system when the black holes
are sufficiently separated and the gravitational field is weak. Thus,
a promising approach for
constructing long waveform models covering the inspiral, merger and
ringdown regimes is to stitch together the results of PN and NR
calculations. 

One procedure for constructing such hybrid waveforms is presented in
\cite{Ajith:2007qp,Ajith:2007kx,Ajith:2007xh}, where
PN and NR waveforms are matched in an appropriate regime 
($-750\,\le\,t/M\,\le\,-550$) prior to the
merger ($M$ is the total mass of the binary system in solar
masses). Restricted 3.5PN waveforms at mass-quadrupole
order are used for the inspiral phase, as given by equation (3.1)
of~\cite{Ajith:2007kx}. For the numerical part, the model is based on
long unequal-mass 
waveforms from simulations run by the Jena 
group using the \texttt{BAM}
code~\cite{Bruegmann:2006at,Hannam:2007ik,Damour:2008te}. 
These simulations span a range of mass ratios corresponding to $0.16
\le \eta \le 0.25$, where $\eta=(m_1 m_2)/M^2$ is the so-called
symmetric mass ratio of the binary system. The matching of PN and NR
data is performed over an overlapping region, under the
assumption that both
approaches to the true BBH waveform are approximately correct at the
late inspiral stage. Once 
the hybrid waveforms are constructed, they are fit to a
phenomenological model
determined entirely by the physical parameters of the binary
system. This fit
to an analytical expression is performed in the Fourier domain,
assuming a functional dependence of the form
\begin{equation}
\label{phen_ansatz} 
u(f) = A_{\mathrm{eff}}(f)\rme^{\rmi\Psi_{\mathrm{eff}}(f)},
\end{equation}
with amplitude and phase given by equations (11) and (14)
of~\cite{Ajith:2007kx}. Each waveform is parametrized by the physical 
parameters of the system, which in the non-spinning case are solely the
masses $m_1$ and $m_2$ of the black holes. As a result of the matching
and fitting procedures described above, a
two-dimensional template family of waveforms that attempt to model
the entire coalescence of non-spinning binary black
hole systems has been obtained.

\begin{figure}
  \includegraphics[width=0.52\textwidth]{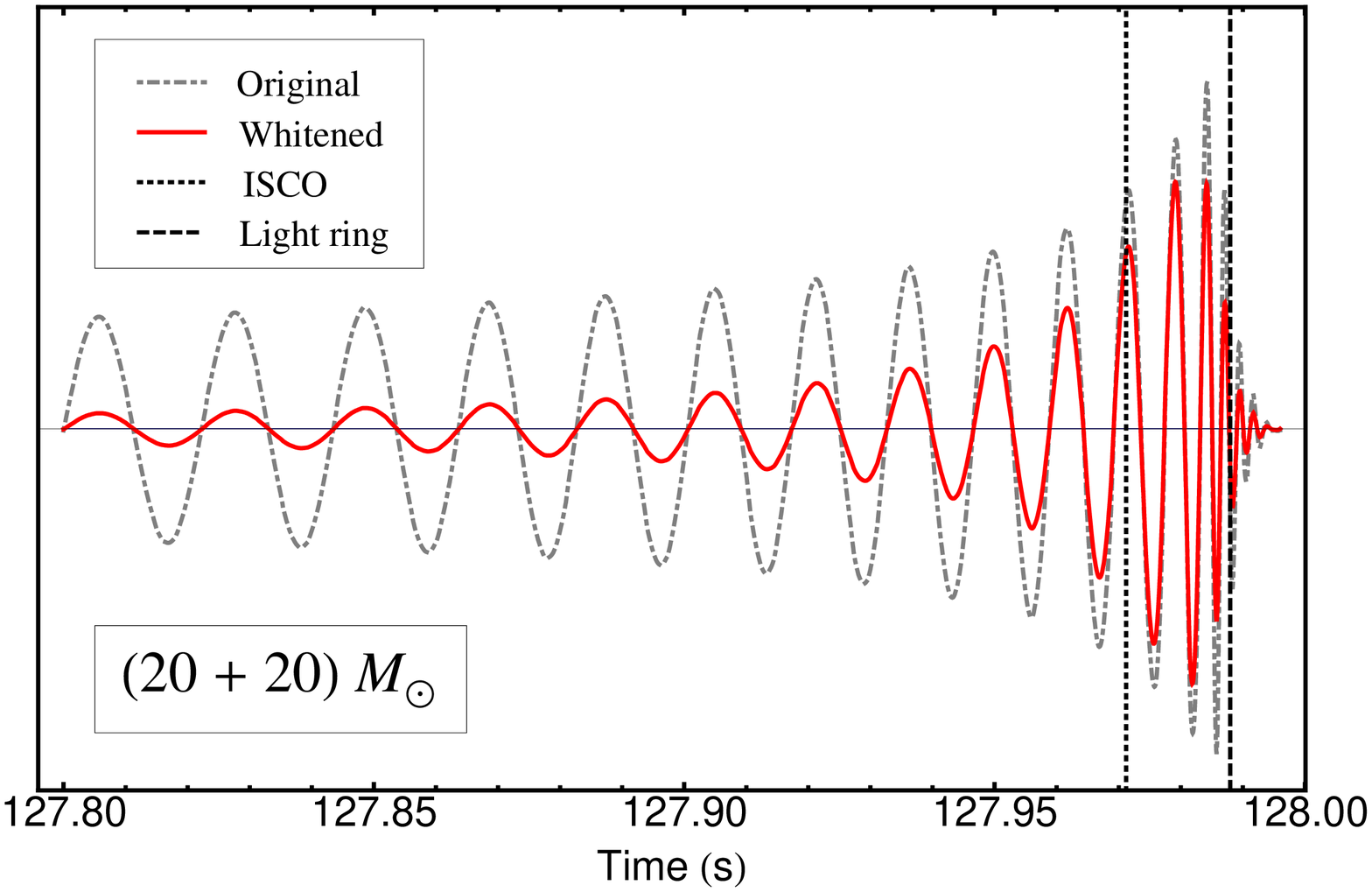}
  \includegraphics[width=0.52\textwidth]{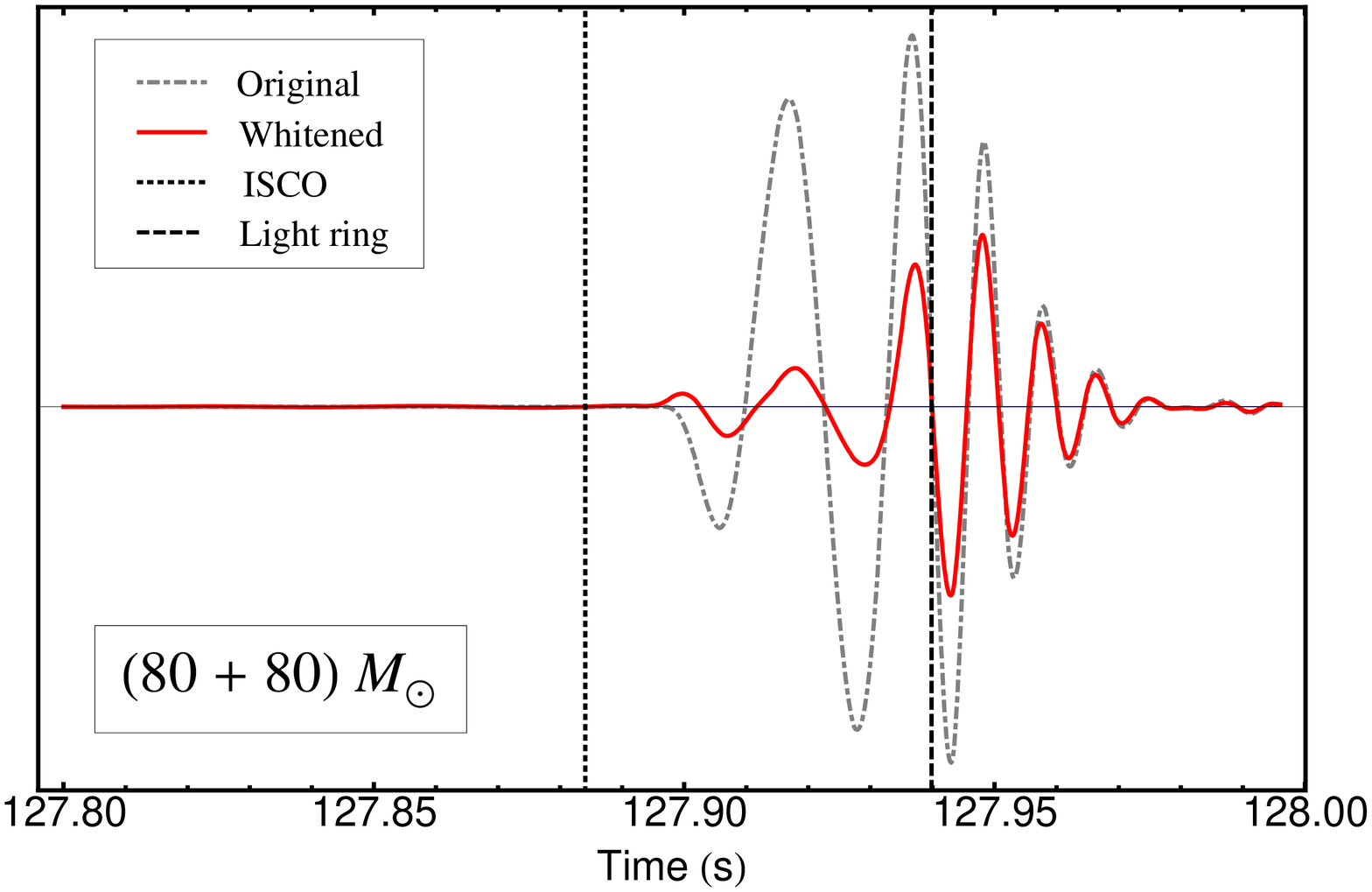}
  \caption{The figure shows two time-domain phenomenological waveforms
    from the template bank used in this search, corresponding to
    equal-mass binaries in the corners of our parameter space, namely
    $(20+20)M_\odot$ and $(80+80)M_\odot$ BBH systems. The
    original and ``whitened''~\cite{Damour:2000gg} waveforms are
    shown, with their
    amplitudes arbitrarily resized. The dotted and dashed vertical
    lines mark the points where the ISCO and light ring frequencies
    are reached. The LRD frequency is not shown, since it basically
    extends up to the full waveform. A matched-filter search that
    starts at 30 Hz and ends at the ISCO will not be able to pick the
    most massive binaries, since the inspiral phase of the coalescence
    falls below the LIGO interferometers' detection band. It is expected that
    the light ring and LRD frequencies, which extend up to the BH
    merger and ringdown respectively, will show improved performance at
    recovering high-mass signals.}
  \label{fig:phenomtmpl}
\end{figure}

The phenomenological template bank has been included in the LSC inspiral
pipeline routinely used by the CBC group as a new waveform for filtering
in the time domain. A search on the NINJA data has
been performed, within a mass range of 
$20 M_{\odot} \leq m_1, m_2 \leq 80 M_\odot$ for the component masses,
with 
$40 M_{\odot} \leq M \leq 160 M_\odot$ for the total mass of the
binary. A technical issue regarding construction of high mass
waveforms prevented us from using a parameter space that completely
overlaps with the mass range of the injected NINJA signals
(see Fig.~\ref{fig:ninjaInjPars}), which might negatively affect
the capability for recovering high-mass signals. Further searches for
prospective NINJA projects will greatly benefit from inclusion of
high-mass templates, and a fix for this issue is underway.
The template bank is constructed using the
standard second order post-Newtonian metric, and uses a hexagonal
placement algorithm in mass space with a minimal match of
0.99~\cite{Cokelaer:2007kx}. Note that we have not implemented the new
metric correspondent to the phenomenological waveforms, which might be
significantly different than the standard 2PN metric; however, as we
shall see, the current template bank placement suffices for detection
purposes and for moderately accurate parameter estimation.
The number of signals that are recovered by the 
pipeline depends strongly on the choice for the upper frequency cutoff
used in the matched filter integral, as we 
have observed in our investigations with the integration stopping at the
ISCO (Innermost Stable Circular Orbit, $r=6M$), 
light ring (the unstable circular orbit 
for photons orbiting a Schwarzschild black hole, $r=3M$) and
Lorentzian ringdown (LRD)\footnote{This is 1.2 
times the fundamental ringdown frequency of Berti, Cardoso and Will, 
Phys. Rev. D \textbf{73} 064030 (2006).}
frequencies.

In Figure~\ref{fig:phenomtmpl} we show two waveforms from our
phenomenological template bank, which correspond to equal-mass binaries
in the corners of our parameter space, namely total
mass $M=40 M_\odot$ and $160 M_\odot$. Displayed are both the original
time domain waveform and its
``whitened'' form~\cite{Damour:2000gg}, as the initial LIGO detector
perceives it, and the
relative amplitudes have been arbitrarily
resized. The whitened waveform is computed as the inverse
Fourier transform of the original signal multiplied by the function 
$1/\sqrt{S_h(f)}$ in the frequency domain, where $S_h(f)$ is the
one-sided noise power
spectral density of the simulated LIGO detectors. In each plot the vertical
lines correspond to the ISCO and light ring frequencies. 
In our searches we have started filtering against the phenomenological
templates at either 30 or 40 Hz and we have stopped the integration at
the three frequencies discussed in the above paragraph. It is evident
that whereas a cut at the ISCO frequency still retains a good portion of the
inspiral signal for low-mass binaries, it is insufficient for higher
masses. The light ring and LRD frequencies, on the other hand, extend roughly
up to the BH merger and to the Lorentzian tail (from the decay of the
quasi-normal modes of the ringdown), respectively, and are therefore
expected to produce more efficient results for a matched-filter search for
high-mass signals.

\section{Results}
\label{sec:results}

\subsection{Efficiency for detection}
\label{sec:detection}
The main results of our search for numerical relativity signals
injected in simulated LIGO noise employing a phenomenological 
template 
bank are presented in Table~\ref{tab:found-missed}. We show here a
summary of the found triggers at different stages of the pipeline for
several runs, with
the starting frequency for the matched-filter integral being either $30$
or $40\un{Hz}$ and the integration stopping at three different
frequencies --ISCO, light ring and LRD-- displayed in ascending order. 
We have separated our results in two sections, according to performance
in recovering the full set of
126 NINJA injections and the reduced set of 67 injections whose total
mass falls below $160 M_\odot$. This choice is motivated by the
construction of the phenomenological bank discussed in \Sref{sec:phenom}.

\begin{table}
\begin{tabular}{| l || c | c | c | c |}
\hline
\bf{Template} & Phenom & Phenom & Phenom & Phenom \\ \hline
\bf{Frequency Cutoff} & ISCO & LightRing & LRD & LRD  \\ \hline
\bf{Filter Start Frequency} & 30 Hz & 30 Hz & 30 Hz & 40 Hz  \\ \hline

\multicolumn{5}{|c|}{Complete set of 126 NINJA Injections} \\ \hline

\bf{Found Single (H1, H2, L1)} & 78, 54, 69 & 94, 66, 90 & 92, 61, 87 & 93, 60, 86 \\ \hline
\bf{Found Coincidence} & 59 & 78 & 81 & 80 \\ \hline
\bf{Found Second Coincidence} & 59 & 80 & 80 & 79 \\ \hline

\multicolumn{5}{|c|}{Reduced set of 67 NINJA Injections with $M < 160 M_\odot$} \\ \hline

\bf{Found Single (H1, H2, L1)} & 40, 17, 32 & 55, 41, 50 & 55, 41, 50
& 56, 40, 50 \\ \hline
\bf{Found Coincidence} & 30 & 47 & 48 & 47 \\ \hline
\bf{Found Second Coincidence} & 30 & 47 & 48 & 47 \\ \hline

\end{tabular}
\caption{Results of the search for NINJA  signals using the 
  phenomenological template bank. There were 126 injections performed into 
  the analyzed data
  for H1, H2 and L1, 67 of which fell within the mass range of our
  phenomenological template bank ($M < 160M_\odot$). We
  explicitly show that a much better efficiency
  in trigger recovery 
  is achieved when the cutoff frequency is pushed beyond the ISCO
  frequency, up to the light ring and Lorentzian ringdown frequencies. 
  Likewise we observe improved efficiency in finding the signals
  that lie within the mass range of our template bank. In both cases
  the signal-based vetoes
  have little influence in the rejection of triggers, confirming their
  efficiency in separating inspiral-like signals from other kind of
  glitches.} 
\label {tab:found-missed}
\end{table}

A time window of $120\un{ms}$ has been used in order to cluster the 
triggers 
found by the pipeline in a single detector. 
Similarly the coincidence has been determined within a $80\un{ms}$ injection 
window. Given these choices for the parameters used in clustering the
triggers, we  
report recovery of 80/126 triggers in double or triple coincidence for
the full injection set and 48/67 triggers for the reduced
set with $M<160M_\odot$.
These are triggers that survive the second coincidence stage (including
the signal-based vetoes) for our best run, which corresponds to the
matched-filter integral starting at $30\un{Hz}$ and ending at the
LRD frequency. The number of recovered triggers in the full mass range
is compatible with
the results quoted by other participants in the NINJA project
employing searches with higher-order corrections PN templates extended
up to larger frequencies, such as the
effective ringdown (ERD) and weighted ringdown\footnote{An
  intermediate  value
  between the ISCO and the ERD frequencies.} (WRD) ending
frequencies~\cite{Aylott08}. The efficiency of the search improves,
however, when we restrict ourselves to signals with masses overlapping
those of our template bank. It is worthwhile noting that among the
triggers recovered by the pipeline we find not only non-spinning
simulations but also signals with non-precessing spins, such as the
\texttt{CCATIE} and \texttt{BAM\_BBH} (except the $a/m=0.25$ case, see below)
waveforms. This supports existing evidence for the fact that
non-spinning templates should be able to detect non-precessing spinning
signals with moderate individual spins. The capability
of non-spinning templates for recovering signals with precession and
large spins could however be compromised, as we discuss below. Due to the
low statistics of the present analysis, these statements should be
taken with the appropriate reservations.

\begin{figure}
  \centering{
  \includegraphics[width=0.8\textwidth]{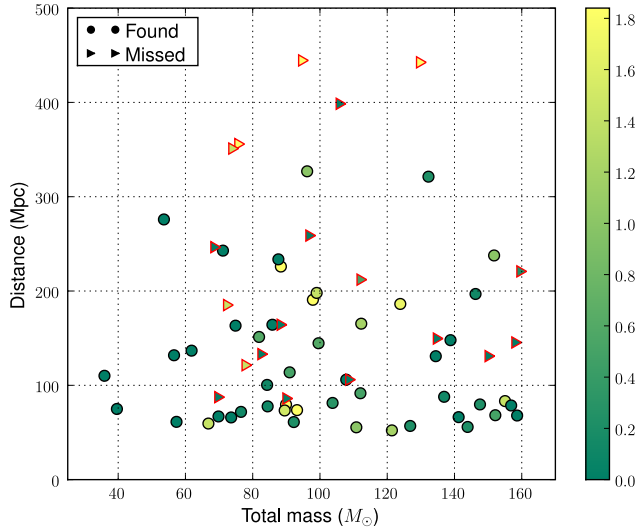}}
  \caption{The figure shows found and missed injections in the mass
    region $32M_\odot\,\le M \,\le 160M_\odot$ as a function
    of their total mass and distance for our
    best search, starting at 30~Hz and stopping at the LRD frequency. The
    circles represent triggers that were recorded as either double or
    triple coincidences after the second stage (including the
    signal-based vetoes), whereas the triangles represent missed
    injections. 
    The colour code represents the modulus of the dimensionless total spin
    $|\vec{S_1}/m_1^2+\vec{S_2}/m_2^2|$ of the black holes.}
  \label{fig:lowMassFoundMissed}
\end{figure}

\begin{table}
\begin{tabular}{| c  c | c  c  c  c  c  c  c |}
\hline
\multirow{2}{*}{ID} & {NR} & Tot. mass & Distance & Eff.~dist. &
Eff.~dist. & $\eta$ & Spin & Tot. Spin \\
 & Simulation & ($M_\odot$) & (Mpc) & H (Mpc) & L (Mpc) & & $a/m$ & $|\vec{S}|$ \\ \hline
136 & \texttt{RIT}  & 94.6  & 444.5 & 15831.9 & 2941.3 & 0.25 & 0.92 & 1.84  \\ 
\textbf{141} & \texttt{RIT}  & 75.6 & 355.8 & 1047.2 & 746.6 & 0.25 &  0.92 & 1.84  \\ 
\textbf{142} & \texttt{RIT} & 129.7 & 442.5 & 2221.8 & 1537.7 & 0.25 & 0.92 &  1.84  \\ 
\textbf{59} & \texttt{CC} & 69.7 & 87.5 & 469.2 & 1573.0 & 0.25 & 0 & 0 \\ 
\textbf{41} & \texttt{BAM\_HHB\_spp25} & 112.0 & 212.1 & 1150.8 & 802.2 & 0.25 & 0.25 & 0.5  \\ 
47 & \texttt{BAM\_HHB\_spp00} & 150.0 & 131.0 & 648.5 & 908.0 & 0.25 & 0 & 0 \\ 
27 & \texttt{BAM\_FAU} & 77.8  & 121.3 & 764.1 & 741.1 & 0.25 & 0.75 & 1.43  \\ 
\textbf{29} & \texttt{BAM\_FAU} & 72.3  & 185.1 & 1325.7 & 885.4 & 0.25 & 0.75 &  1.43  \\ 
\textbf{30} & \texttt{BAM\_FAU} & 73.9  &  351.1 & 896.9 & 771.4 & 0.25 & 0.75 &  1.43  \\ 
114 & \texttt{PU\_T52W} & 82.6  & 133.1 & 364.0 & 320.2 & 0.25 & 0 & 0 \\ 
116 & \texttt{PU\_T52W} & 88.2  & 164.2 & 654.7 & 533.7 & 0.25 & 0 &  0  \\ 
118 & \texttt{PU\_T52W} & 90.0  & 86.0 & 1055.4 & 452.3 & 0.25 & 0 &  0  \\ 
120 & \texttt{PU\_T52W} & 108.6  & 106.0 & 202.0 & 205.9 & 0.25 & 0 &  0  \\ 
\textbf{125} & \texttt{PU\_T52W} & 96.8  & 258.8 & 463.8 & 464.7 & 0.25 & 0 &  0  \\ 
\textbf{126} & \texttt{PU\_T52W} & 105.8  & 398.5 & 1175.2 & 1797.1 & 0.25 & 0 &  0  \\ 
\textbf{64} & \texttt{GSFC\_X4} & 134.7  & 149.6 & 1320.4 & 856.0 & 0.16 & 0 &  0  \\ 
\textbf{68} & \texttt{GSFC\_X3} & 160.0  & 220.8 & 819.3 & 1123.2 & 0.1875 & 0 &  0  \\ 
\textbf{76} & \texttt{GSFC\_X4} & 158.0  & 145.4 & 722.2 & 558.5 & 0.16 & 0 &  0 \\ 
\textbf{95} & \texttt{Lean\_c138} & 68.6  & 246.5 & 333.9 & 407.7 & 0.16 & 0 &  0  \\ \hline
\multicolumn{9}{|c|}{Total number of missed injections: 19}  \\ \hline
\end{tabular}
\caption{Overview of the 19 missed injections with total mass below
  $160M_\odot$ for the best run reported. The ID column stores an
  index that identifies each of the injections of the NINJA set. The
  convention for the naming of the NR simulations can be found
  in~\cite{Aylott08}. Among
  the missed signals we
  stress the presence of waveforms with eccentricity, large spins and 
  precession and
  also those injected at distances further than $350\un{Mpc}$. Note
  that the boldfaced IDs correspond to signals also reported as missed
  in~\cite{Aylott08aa}, where a Bayesian inference search on the NINJA
  data using a Nested
  Sampling algorithm is presented. }
\label{tab:missedInj}
\end{table}

Figure~\ref{fig:lowMassFoundMissed} provides an overview of the found
and missed injections corresponding to total mass below
$160M_\odot$. The colour code encodes the modulus of the dimensionless
total spin
$|\vec{S_1}/m_1^2+\vec{S_2}/m_2^2|$ of the black holes, and gives an
indication of the injections that significantly deviate from the
non-spinning case 
modelled by the phenomenological waveforms. We observe how
signals located at distances above $350\un{Mpc}$ are systematically
lost, giving us an indication of the distance reach of the pipeline;
nevertheless, several nearby injections are
missed as well. In order to track down the missed injections in the
mass region below $160M_\odot$, a compilation of 
their relevant physical parameters and associated information is given
in Table~\ref{tab:missedInj}. For a description of the diverse NR
simulations listed therein we refer the reader
to~\cite{Aylott08}. A similar analysis of the missed and found
injections has been recently performed by the Birmingham
group in~\cite{Aylott08aa}, applying Bayesian inference on the NINJA
data using a Nested Sampling algorithm. The work of Aylott~\etal
explores how different waveform
families affect the confidence of detection of NR waveforms. Their
Bayes factor $B$ is a metric for
assessing the level of confidence that a signal has been detected, and
their defined thresholds for $\log_{10}B$ allow for classification of the
signals as found or missed.
The IDs displayed in bold type in our Table~\ref{tab:missedInj}
correspond to
signals that are reported as missed by the Birmingham group in Table~2
of~\cite{Aylott08aa}; 12 of our 19 missed injections are also lost by
them, a correlation that seems worth following up. Future versions of
the NINJA
project will certainly benefit from combined searches and
cross-checks of this kind between different DA methods. 

Signals with large eccentricity, such 
as the Princeton $e \gtrsim 0.5$ \texttt{PU\_T52W} run are invariably 
lost by our pipeline. Likewise, the phenomenological templates are not
able to pick up
signals with considerable spin, such as the equal-mass, spinning
waveforms from \texttt{RIT} with individual spins $a/m=0.92$ pointing
along the z-axis and \texttt{BAM\_FAU} with randomly aligned spins
$a/m=0.75$. Our template
bank was developed to search for signals
in which spin is unimportant and no precession is present, so these
results are understandable. 
Further work targeted to incorporating spins within the 
phenomenological model is desirable and will be undertaken
in the future. The missed
\texttt{GSFC} signals with mass ratio $1:3$ and $1:4$ correspond to
runs with few orbits before merger and moreover they are injected at
total masses bordering on the edge of our template bank, which could
explain them being lost. More bewildering is however
the fact that the pipeline misses a couple of long equal-mass
non-spinning simulations injected at
rather close distances, such as \texttt{CaltechCornell}, which is also
missed by the search reported in~\cite{Aylott08aa}, and
\texttt{BAM\_HHB\_spp00}. A look at the columns of
Table~\ref{tab:missedInj} that list the \textit{effective distance} in
H and L (the distance to an equivalent source with optimal location and
orientation) indicates that it might be the poor orientation of these injections
that prevents the pipeline from finding them.
Aside from these individual cases, which would
need a careful follow-up that is below the scope of this paper, we
can justify the rest of missed injections as those either placed at
large distances, presenting large spin values and/or precession
and containing few orbits before merger. 

Among the obvious improvements that a search with phenomenological
templates could benefit from we can mention the following. Firstly, and once the
technical issue with the generation of high-mass templates is resolved,
the search would clearly improve with the use of a template bank that
fully 
covers the parameter space of the signals searched for. Additionally,
the inclusion of the fourth interferometer V1 in our pipeline shall
provide a larger number of recovered triggers, in the manner reported
by the search using
EOBNR templates that is described in Section 4.1.3
of~\cite{Aylott08}. Both improvements will be most likely incorporated to
searches with the phenomenological template bank in future
realizations of the NINJA project.

\subsection{Efficiency for parameter estimation}
\label{sec:parEstimation}

\begin{figure}
\includegraphics[width=0.6\textwidth]{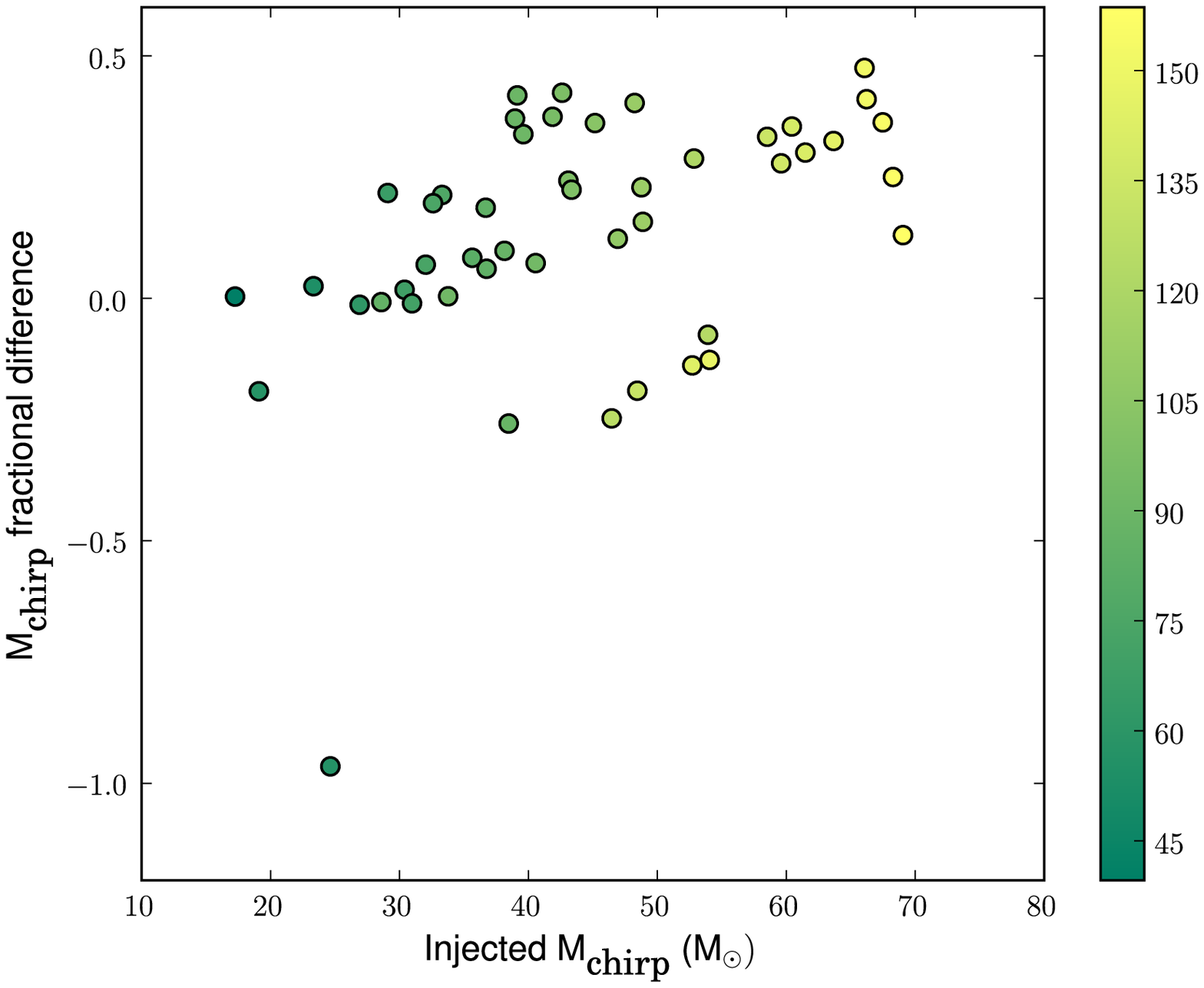}
\includegraphics[width=0.6\textwidth]{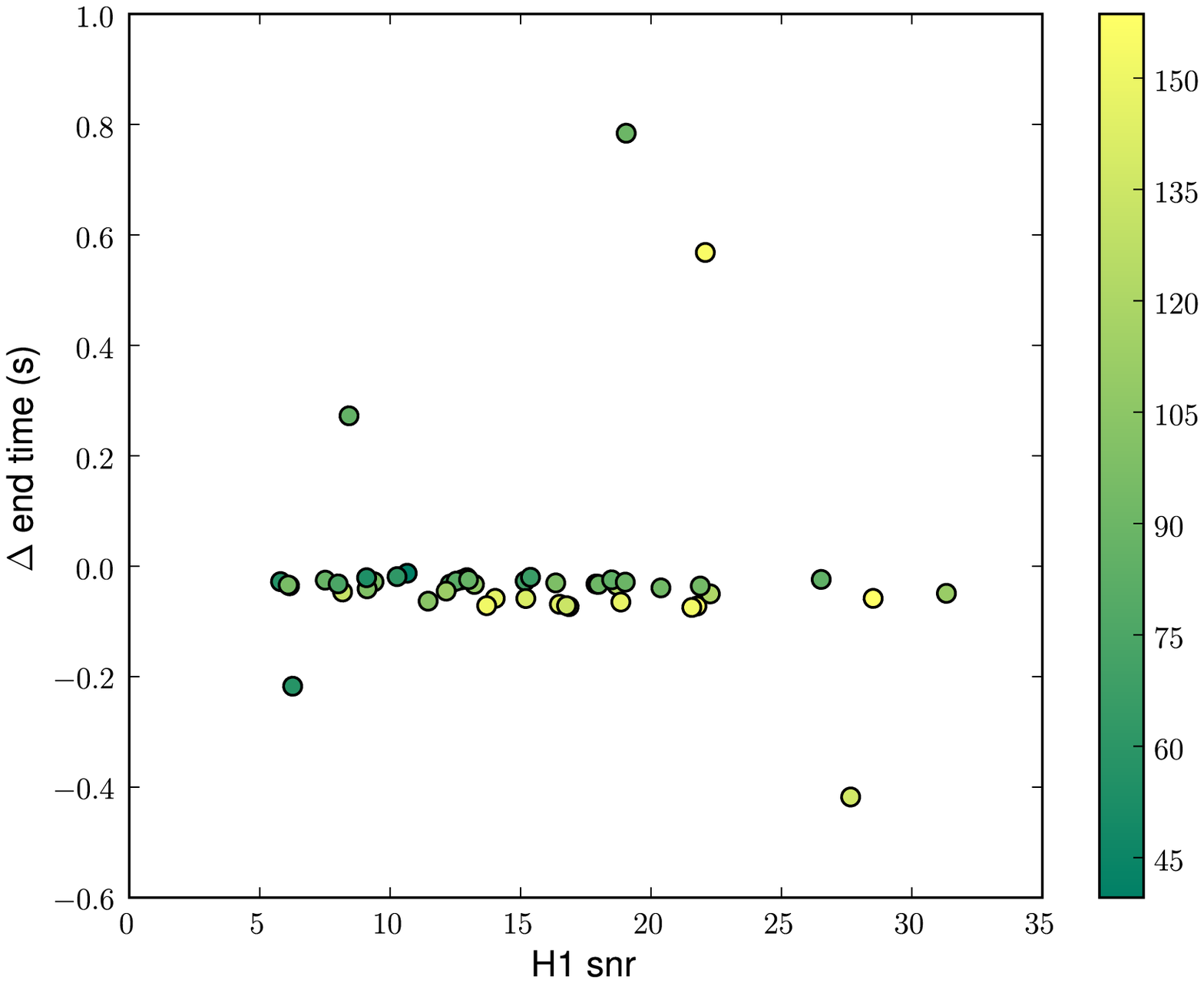}
\caption{Figures showing the accuracy with which the chirp mass (left
    panel) and end time (right panel) of the reduced set of NINJA injections
    with total mass below $160M_\odot$ are recovered using the
    phenomenological template bank. In both plots,
    the colour
    scale is given by the total mass of the system. The chirp mass,
    a common quantity in data analysis which is defined as
    $\mathcal{M} = (m_1 m_2)^{3/5} (M)^{-1/5}$, is 
    typically recovered within a $20\%-40\%$ accuracy, depending on the
    chirp mass and total mass of the system. On the right panel we
    observe a few low-mass outliers
    that would need a more careful follow-up. Nevertheless, the
    results for parameter estimation with our IMR bank constitute a
    significant improvement over current LIGO/Virgo searches with standard PN
    templates. }
\label{fig:PhenomParam} 
\end{figure}

The number of found versus missed triggers is not the only relevant
metric for assessing the performance of the standard GW searches. If
astrophysically relevant statements are to be made from GW
observations, the ability of accurately estimate the physical
parameters of the measured signals is
crucial. Figure~\ref{fig:PhenomParam} shows parameter estimation for
the phenomenological search on the NINJA injections. The left panel
displays the fractional difference for recovery of the chirp mass of the 
system, defined as $\mathcal{M} = (m_1 m_2)^{3/5} (M)^{-1/5}$. The
vertical colour bar encodes the total mass in solar masses.
We report substantial improvement in parameter estimation with
respect to LIGO/Virgo standard searches that make use of PN templates
(see Figure 8 of the NINJA paper~\cite{Aylott08}), although it should
not be forgotten that the PN searches reported in~\cite{Aylott08} make
use of banks with masses up to $90M_{\odot}$ only. In our
case the chirp mass is recovered within a $20\%$ accuracy for values
below $40M_\odot$ and $\sim40\%$ for signals with larger chirp mass.
The outlier that can be
spotted at $\mathcal{M}\sim 30 M_\odot$ corresponds to a
\texttt{CaltechCornell} waveform injected at $132\un{Mpc}$ with total
mass $56.6M_\odot$. For this particular injection the accuracy in
parameter recovery is rather poor, and further work to understand this
behaviour will be undertaken in the future.

The panel in the right shows the accuracy in end time recovery of the
found signals, with the 
colour code again displaying the total mass of the system. The sign
convention for the $\Delta t_{\mathrm{end}}$ corresponds to the
injected minus 
the recovered parameters, so that a trigger that presents a positive
value $\Delta t_{\mathrm{end}} > 0$ indicates that the signal
was really injected at a \textit{later} time than the value recorded
by our pipeline. Most of the signals are recovered at a time within a
few hundredths of second from the injected end time value, with the
outliers corresponding partially to
signals with larger total mass. A more careful study of the meaning of
the end time when we are dealing with NR signals is needed in order to
improve accuracy in estimation of this parameter. 

Even though the number of total recovered triggers for the
phenomenological search on the NINJA data is similar to the results
quoted by the standard PN searches, it is clear that the use of a full
IMR 
template bank helps the estimation of the physical signal
parameters. In view of these and other coincident
results quoted 
in~\cite{Aylott08,Farr08,Aylott08aa}, we conclude that
searches that attempt to recover and estimate the physical
parameters of BBH signals in the mass
range $10^2 - 10^3 M_\odot$ would profit from using
an IMR template bank that fully models the inspiral, merger and ringdown of the
binary system. This is of crucial important
for future LIGO/Virgo searches that aim at targeting coalescences of
compact objects in the above-mentioned range, for which a full
template bank adapted to arbitrarily high masses (and
ideally also to non-zero spin values) needs to be developed. Attempts
in this direction are already underway within the LIGO/Virgo
collaboration.

\section{Conclusion}
\label{sec:conclusion}

We have presented the results from the first search using the standard
inspiral 
pipeline modified to match-filter against the  
phenomenological template bank introduced by Ajith \textit{et al}, an
inspiral-merger-ringdown template that models the full coalescence of
non-spinning black hole binaries of small mass ratios. This
procedure has been directly applied to the search for
numerically-simulated signals injected into simulated LIGO noise within
the frame of the NINJA collaboration. We have tried several values for
the cutoff frequency in the match-filter integral and found results
that corroborate the need for pushing the integration to higher
values than the ISCO
if one wants to pick up signals that contain information about the
merger and ringdown stages of the BBH coalescence process, which is
very relevant for LIGO/Virgo high-mass searches. The total
number of NINJA signals that we are able to recover using the
phenomenological template bank is comparable to the results reported
by other groups that participated in the NINJA project searching with PN
templates extended to higher frequencies. Nevertheless, improved
accuracy in parameter recovery is obtained with our phenomenological
template bank. Moreover, it is expected that, had we constructed a
bank covering the exact parameter space of the NINJA injections and
included the fourth V1 detector in our analysis, the
number of recovered triggers would have significantly increased. This
information will be used as starting point in future applications of the
phenomenological waveforms to the search for inspiral signals.

There is however a number of signals that are not found by
our pipeline and
that we have identified as injections falling outside the
parameter space of our template bank, injections placed at distant
locations above $350\un{Mpc}$, poorly oriented injections and
injections corresponding to evolutions 
of black holes with significant eccentricity, spins and precession,
plus a few outliers that undoubtedly need a more careful and dedicated
study. More precise statements on found and missed signals would
nevertheless require larger statistics in our data.

There are several obvious improvements in the construction of the
phenomenological waveform model that can be made. The phenomenological
waveforms have so far been
constructed only for the dominant mode of the comparable mass
non-spinning case and it is not surprising that, for example, some
waveforms with precession or eccentricity are not recovered. 
Searches for spinning objects are of the greatest
astrophysical importance, and phenomenological or other types of
theoretical models for these kind of systems are much needed.
The phenomenological ansatz used to fit
the merger portion (especially the power-law for the amplitude) needs
to be extended to higher mass ratios and spins. The error bars on the
phenomenological parameters need to be better quantified, and the
matching to post-Newtonian theory done as early in the inspiral phase
as possible, and higher modes included. Notwithstanding, 
our results show a clear improvement in parameter estimation
when the full inspiral-merger-rindown waveform is used as
template. This result is confirmed by other reports within the frame
of the NINJA collaboration. Future research on phenomenological models
comprises development of adequate phenomenological waveforms
targeted at getting better detection and parameter estimation of signals
from black-hole binaries in the current LIGO/Virgo data-analysis
pipelines. Work
on these improvements is in progress, and will be applied in future
NINJA projects.

\section*{Acknowledgements}
The authors thank S.~Fairhurst, B.~Farr, E.~Ochsner, 
B.~Satyaprakash, M.~Hannam, S.~Husa and L.~Pekowski for useful
comments and helpful 
suggestions while this analysis was being carried out. 
We acknowledge hospitality from the Kavli Institute for Theoretical
Physics (KITP) Santa Barbara during the workshop ``Interplay between
Numerical Relativity and Data
Analysis'', where the NINJA project was initiated; the Kavli Institute
is supported by NSF grant PHY 05-51164.
This work was supported by the Max Planck Society, by DFG grant
SFB/TR~7, by the German Aerospace Center (DLR), and by the
National Science Foundation under grant NSF-0838740. LS is partially
supported by DAAD grant A/06/12630. This paper has been assigned LIGO
document number P080125-01-Z.

\section*{References}

\bibliographystyle{iopart-num}

\bibliography{NRDA08refs}

\end{document}